\documentclass[aps,pre,superscriptaddress,twocolumn,floatfix] {revtex4}
\usepackage{times}
\usepackage{graphicx}
\usepackage[dvips,unicode,colorlinks,linkcolor=blue,citecolor=blue,urlcolor=blue]{hyperref}

\begin{document}
\title{
The COMPASS force field: validation for carbon nanoribbons.
}
\author{A. V. Savin}
\email[]{asavin@center.chph.ras.ru} 
\affiliation{Semenov Institute of Chemical Physics, 
Russian Academy of Sciences, ul. Kosygina 4,
119991 Moscow, Russia}
\author{M. A. Mazo}
\email[]{mikhail.mazo1@gmail.com} 
\affiliation{Semenov Institute of Chemical Physics, 
Russian Academy of Sciences, ul. Kosygina 4,
119991 Moscow, Russia}

\begin{abstract}
The COMPASS force field has been successfully applied in a large number of materials simulations, 
including the analysis of structural, electrical, thermal, and mechanical properties of carbon 
nanoparticles. This force field has been parameterized using quantum mechanical data and 
is based on hundreds of molecules as a training set, but such analysis for graphene sheets was not 
carried out. The objective of the present study is the verification of how good the COMPASS 
force field parameters can accurately describe the frequency spectrum of atomic vibrations 
of graphene, graphane and fluorographene sheets. We showed that the COMPASS force field allows 
to describe with good accuracy the frequency spectrum of atomic vibrations of graphane and 
fluorographene sheets, whose honeycomb hexagonal lattice is formed by sp$^3$ hybridization. 
On the other hand, the force field doesn't describe very well the frequency spectrum of graphene 
sheet, whose planar hexagonal lattice is formed by sp$^2$ banding. In that case the frequency 
spectrum of out-of-plane vibrations differs greatly from the experimental data -- 
bending stiffness of a graphene sheet is strongly over estimated. We present the correction 
of parameters of out-of-plane and torsional potentials of the force field, that allows to achieve 
the coincidence of vibration frequency with experimental data. After such corrections 
the COMPASS force field can be used to describe the dynamics of flat graphene sheets 
and carbon nanotubes.
\end{abstract}


\maketitle

\section{Introduction\label{sc1}}

The obtaining of the monolayer graphene membrane with the unique physical properties \cite{pm1}
caused the unprecedented rise in research of single and multilayer graphene sheets, 
graphene nanoribbons and nanoscrolls, other graphene-based and functional graphene nanostructures
\cite{pm2,pm3,pm4,pm5,pm6,pm7,pm8}. The possibility of using such structures in electronics
\cite{pm9,pm10}, optics \cite{pm11}, and in many other industries has been much discussed recently 
\cite{pm5,pm7,pm12,pm13,pm14,pm15,pm16}. The remarkable properties of graphene make it one 
of the key components in creating nanomaterials for energy storage \cite{pm17,pm18,pm10,pm20}, 
polymer nano\-composites \cite{pm2,pm21} and for medicine \cite{pm22,pm23}.

Experimental studies of these nanostructures are difficult because of their small size. 
So particular attention is paid to their computer simulation, where quantum mechanics (QM) 
equations, molecular dynamics (MD) or molecular mechanics (MM) equations are used. 
QM simulation allows to consider only small-sized molecular structures, 
but enables to justify empirical interatomic potentials in MM/MD modelling. 
However, the accuracy of MD simulations depends on parametrization of the empirical potentials 
that describe the atomic interactions. Both experimental data and results of QM calculation 
used for determination of these parameters. For MM/MD modelling carbon nanoforms a variety 
of carbon interatomic potentials have been used. Examples of the carbon potentials are reactive 
empirical bond-order potential REBO/AIREBO \cite{pm24,pm25,pm26,pm27}, 
reactive force field ReaxFF \cite{pm28}, long-range carbon bond order potential LCBOP \cite{pm29}, 
an analytical bond order potential (BOP) \cite{pm30}, environment dependent interatomic 
potential (EDIP) \cite{pm31}, the modified embedded atom method (MEAM) potential \cite{pm32}, 
the DREIDING force field \cite{pm33}, Morse force field \cite{pm34,pm35}. 

It is assumed that the force field parameters should reproduce the value of mechanical moduli 
and vibration spectrum of nanoparticles, but the standard parameter sets of the above-mentioned 
force fields have been received taking into account only in-plane frequency spectrum or/and 
in-plane molecular mechanics. Taking into account out-of-plane vibrations made it possible 
to significantly improve the matching of the bending rigidity modulus and dispersion curves 
calculated in MD with the available experimental data and QM simulation for force fields 
Morse \cite{pm36,pm37}, MEAM \cite{pm38} and DREIDING \cite{pm39}.
\begin{figure}[t]
\begin{center}
\includegraphics[angle=0, width=1.0\linewidth]{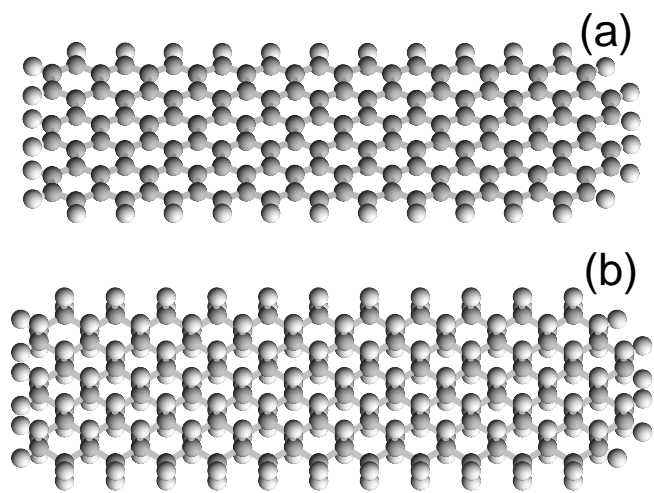}
\end{center}
\caption{
Structures of nanoribbons (a) graphene  (C$_{12}$H$_2$)$_{11}$C$_{10}$H$_{12}$  
(of size $29.9\times 13.4$\AA$^2$)
and (b) graphane  (C$_{12}$H$_{14}$)$_{11}$C$_{10}$H$_{22}$ (of size $29.1\times 11.7$\AA$^2$).
}
\label{fig1}
\end{figure}
\begin{figure}[t]
\begin{center}
\includegraphics[angle=0, width=1.0\linewidth]{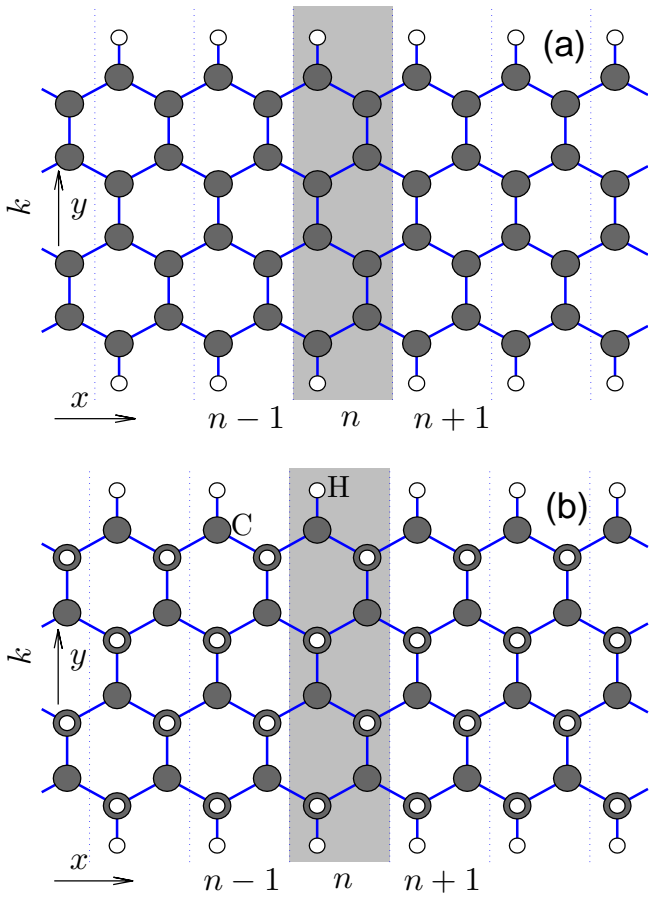}
\end{center}
\caption{
Schematic view of the structures of zigzag (a) graphene (C$_8$H$_2$)$_\infty$ 
and (b) graphane (C$_8$H$_{10}$)$_\infty$ nanoribbons. 
The indices $n$ and $k$ enumerate the unit cells in the nanoribbon
and the atoms in the cell. Dotted lines indicate the cell borders.
Gray color region is the elementary cell of the nanoribbon.
Nanoribbons lie on the $xy$-plane.
}
\label{fig2}
\end{figure}

A condensed-phase optimised ab-initio COMPASS force field has been developed recently
\cite{pm40,pm41} and has been successfully applied in a large number of soft materials simulations 
with carbon nanostructures (see \cite{pm42,pm43,pm44,pm45,pm46,pm47,pm48,pm49} 
and references in them). The force field has been parameterized using hundreds of molecules 
as a training set including molecules with sp$^3$- and sp$^2$-hybridized carbon atoms, 
but the parameterization and validation using planner monomolecular lattice structures 
as graphene sheets and its modifications have not been yet performed.

The purpose of article is to checking how well force field of the COMPASS reproduces 
a vibration range of three nanoribbons: graphene, graphane and fluorographene. 
For this we calculated the frequency spectrum and dispersion curves of these nanoribbons 
using the COMPASS force field, and compared them with results of ab-initio data. 
It was found that the force field could give an accurate reproduction of experimental spectrum 
lattice of sp$^3$ hybridization, such as graphene or fluorographene. However, 
the obtained frequency spectrum of graphene turned out to be noticeably shifted 
to the high-frequency region with respect to the known theoretical and experimental data 
for out-of-plane vibrations of carbon. We made a correction of parameters of torsional potentials, 
what allowed to rectify that deficiency of the COMPASS force field. 
\begin{table*}[tbh]
\caption{
Values of the parameters for valence bond potential
$V_1(b)=k_{b,2}(b-b_0)^2+k_{b,3}(b-b_0)^3+k_{b,4}(b-b_0)^4$,
angle potential
$V_2(\theta)=k_{\theta,2}(\theta-\theta_0)^2+k_{\theta,3}(\theta-\theta_0)^3+k_{\theta,4}(\theta-\theta_0)^4$
and torsional potential
$V_t(\phi)=k_{\phi,1}(1-\cos\phi)+k_{\phi,2}(1-\cos2\phi)+k_{\phi,3}(1-\cos3\phi)$.
}
\label{tab1}
\centering\noindent
{
\begin{tabular}{lcccc}
\hline
\hline
Bond & ~~$b_0$ (\AA)~~ & $k_{b,2}$ (kcal/mol$\cdot$\AA$^2$)~~& 
~~$k_{b,3}$ (kcal/mol$\cdot$\AA$^3$)~~ 
& ~~$k_{b,4}$ (kcal/mol$\cdot$\AA$^4$)\\  
\hline
CC (sp$^2$ atoms) & 1.4170 & 470.8361 & -627.6179 & 1327.6345 \\
CH (sp$^2$ atoms) & 1.0982 & 372.8251 & -803.4526 & 894.3173 \\ 
CC (sp$^3$ atoms) & 1.530 &  299.670 & -501.77 & 679.81 \\
CH (sp$^3$ atoms) & 1.101 &  345.000 & -691.89 & 844.60 \\
CF (sp$^3$ atoms) & 1.390 &  403.032 &   0     & 0\\ 
\hline
Angle & ~~$\theta_0$ (deg)~~ & ~~$k_{\theta,2}$ (kcal/mol$\cdot$rad$^{2}$)~~ &  ~~$k_{\theta,3}$ (kcal/mol$\cdot$rad$^{3}$)~~
& ~~$k_{\theta,4}$ (kcal/mol$\cdot$rad$^{4}$)\\
\hline
CCC (sp$^2$ atoms) & 118.90  & 61.0226 & -34.9931 & 0 \\
CCH (sp$^2$ atoms) & 117.94  & 35.1558 & -12.4682 & 0 \\ 
CCC (sp$^3$ atoms) & 112.67  & 39.5160 & ~~-7.4430 & -9.5583\\
CCH (sp$^3$ atoms) & 110.77  & 41.4530 & -10.6040 & ~5.1290\\
HCH (sp$^3$ atoms) & 107.66  & 39.6410 & -12.9210 & -2.4318 \\
CCF (sp$^3$ atoms) & 109.20  & 68.3715 & 0 & 0 \\ 
FCF (sp$^3$ atoms) & 109.10  & 71.9700 & 0 & 0 \\ 
\hline
Torsion &  ~~$k_{\phi,1}$ (kcal/mol)~~ & ~~$k_{\phi,2}$ (kcal/mol)~~ & ~~$k_{\phi,3}$ (kcal/mol)\\
\hline
CCCC (sp$^2$ atoms)    & 8.3667 & 1.2000 & 0  & \\
CCCH (sp$^2$ atoms)    &  0     & 3.9661 & 0  &\\ 
HCCH (sp$^2$ atoms)    &  0     & 2.3500 & 0 &\\
CCCC (sp$^3$ atoms)    &   0    & 0.0514 & -0.1430 &\\
CCCH (sp$^3$ atoms)    &   0    & 0.0316 & -0.1681 &\\
CCCF (sp$^3$ atoms)    &   0    & 0      & ~0.1500 &\\
HCCH (sp$^3$ atoms)    &  -0.1432 & 0.0617 & -0.1530 &\\
FCCF (sp$^3$ atoms)    &   0 & 0 & -0.1 &\\ 
\hline
\hline
\end{tabular}
}
\end{table*}

\section{A model of a graphene and graphane nanoribbons}

Molecular nanoribbon is a narrow, straight-edged strip, cut from a single-layered molecular plane. 
The simplest example of such molecular plane is a graphene sheet (isolated monolayer of carbon 
atoms of crystalline graphite) and its various chemical modifications: gra\-phane (fully 
hydrogenated on both sides graphene sheet) and fluorographene (fluorinated graphene).
As is known, gra\-phene and its modifications are elastically isotropic materials, 
the longitudinal and flexural rigidity of which is weakly dependent on chirality of the structure. 
Therefore, for definiteness, we will consider nanoribbons with the zigzag structure shown 
in Fig.~\ref{fig1} and \ref{fig2}.

Let us consider a rectangular ribbon cut from a flat sheet
of graphene [Fig.~\ref{fig1} (a)] and graphane [Fig.~\ref{fig1} (b)]
in the zigzag directions.
The nanoribbons structure can be obtained by longitudinal translation of a transverse unit cell 
consisting of $N_e=N_{\rm C}+N_{\rm H}$ atoms (see Fig.~\ref{fig2}).
Number of the carbon atoms in the unit cell is always multiple of two, 
and number of hydrogen atoms $N_{\rm H}=2$ for graphene nanoribbon and 
$N_{\rm H}=N_{\rm C}+2$ for graphane.  
Further we use the following notation (see Fig.~\ref{fig2}): 
each atom is numbered with at two-component index
$\alpha=(n, k)$, where $n=0,\pm1,\pm2,...$ defines the unit cell number
and $k=1,...,N_e$ numbers atoms in the unit cell. 

For numerical simulation we will use COMPASS force-field functional form \cite{pm40}: 
\begin{eqnarray}
E_{total}=\sum_b[k_{b,2}(b-b_0)^2+k_{b,3}(b-b_0)^3+k_{b,4}(b-b_0)^4]\nonumber\\
+\sum_\theta[k_{\theta,2}(\theta-\theta_0)^2+k_{\theta,3}(\theta-\theta_0)^3+k_{\theta,4}(\theta-\theta_0)^4]
\nonumber\\
+\sum_\phi[k_{\phi,1}(1-\cos\phi)+k_{\phi,2}(1-\cos2\phi)\nonumber\\
+k_{\phi,3}(1-\cos3\phi)]\nonumber\\
+\sum_\chi k_{\chi}\chi^2+\sum_{bb'}k_{bb'}(b-b_0)(b'-b_0')\nonumber\\
+\sum_{b,\theta}k_{b\theta}(b-b_0)(\theta-\theta_0)
+\sum_{\theta,\theta'}k_{\theta\theta'}(\theta-\theta_0)(\theta'-\theta_0')\nonumber\\
+\sum_{b,\phi}(b-b_0)[k_{b\phi,1}\cos\phi+k_{b\phi,2}\cos2\phi+k_{b\phi,3}\cos3\phi]\nonumber\\
+\sum_{\theta,\phi}(\theta-\theta_0)[k_{\theta\phi,1}\cos\phi+k_{\theta\phi,2}\cos2\phi+k_{\theta\phi,3}\cos3\phi]\nonumber\\
+\sum_{\theta,\theta',\phi}k_{\theta\theta'\phi}(\theta-\theta_0)(\theta'-\theta_0')\cos\phi\nonumber\\
+\sum_{i,j}q_iq_j/r_{ij}+\sum_{i,j}\epsilon_{ij}[2(r_{ij}^0/r_{ij})^9-3(r_{ij}^0/r_{ij})^6].~~~~~~
\label{f1}
\end{eqnarray}
The functions could be divided into two categories: (1) valence terms including diagonal and
off-diagonal cross-coupling terms and (2) nonbonded interactions terms. The valence terms represent
internal coordinates of bond $b$, angle $\theta$, torsion angle $\phi$, and out-of-plane angle $\chi$,
and  the cross-coupling terms include combinations of two or three internal coordinates.
The nonbond interactions include a LJ-9-6 potentials for the van der Waals (vdW) term and a 
Coulombic potential for an electrostatic interaction.

Graphene has a flat form due to the sp$^2$ hybridization of all valence bonds 
[Fig.~\ref{fig1} (a)]. 
Graphane (fully hydrogenated graphene) as a two-dimensional crystal was predicted in 
theoretical works \cite{pm50,pm51} and was confirmed by experimental data in \cite{pm52}. 
Graphane is an analogue of graphene with the unique properties \cite{pm53,pm54,pm55,pm56}. 
Graphane nanoribbon is a strip with a constant width cut from a two-sized hydrogenated graphene 
sheet [see Fig.~\ref{fig1} (b), \ref{fig2} (b)]. 
Several conformations of the graphane sheet are existed, 
and we consider the most energy-efficient configuration armchair, where hydrogen atoms are 
connected from different sides to all adjacent carbon atoms. Substitution of hydrogen atoms 
for fluorine atoms leads to formation of the fluorographene sheet structure (fully fluorinated graphene).
In comparison with graphene, graphane and fluorographene are not two-dimensional crystal structures,
but corrugated sheets with sp$^3$ hybridized carbon atoms and with atoms of hydrogen or fluorine 
connected to the different sides of the sheet.

For graphene nanoribbon simulation the potentials (\ref{f1}) with the parameters of atom with sp$^2$
hybridization should be used, and for graphane and fluorographene nanoribbon simulation -- 
potentials with the parameters of atom with sp$^3$ hybridization. There quired parameters 
for interatomic interaction potentials are shown in Tabl.~\ref{tab1}. 
Out-of-plane angle potential $V_3(\chi)=k_{\chi}\chi^2$ appears in the Hamiltonian only 
for graphene nanoribbons with planar sp$^2$ valence bands: for C--C--C--C atom group 
$\kappa_\chi=7.1794$~kcal/mol, for C--C--C--H atoms $\kappa_\chi=4.8912$~kcal/mol.
Parameters for cross-coupling interactions potentials are shown in Tables \ref{tab2} and \ref{tab3}.
Parameters for nonvalence interactions can be  found in Table \ref{tab4}.
Parameters values for C and H atoms are taken from \cite{pm40}, 
for fluorine atom the parameters values -- from force field CFF91. 
\begin{table}[tbh]
\caption{
Values of parameters for cross-coupling bond/bond potential
$V(b,b')=k_{bb'}(b-b_0)(b'-b_0')$, bond/angle potential
$V(b,\theta)=k_{b\theta}(b-b_0)(\theta-\theta_0)$
angle/angle potential,
$V(\theta,\theta')=k_{\theta\theta'}(\theta-\theta_0)(\theta'-\theta_0')$
and angle/angle/torsion potential
$V(\theta,\theta',\phi)=k_{\theta\theta'\phi}(\theta-\theta_0)(\theta'-\theta_0')\cos\phi$
}
\label{tab2}
\centering\noindent
{
\begin{tabular}{lc}
\hline
\hline
bond/bond & $k_{bb'}$ (kcal/mol$\cdot$\AA$^2$)\\
\hline
CC/CC (first neighboring, sp$^2$)  & 68.2856 \\
CC/CH (first neighboring, sp$^2$)  & 1.0795 \\ 
CC/CC (second neighboring, sp$^2$) & 53.0 \\
CC/CH (second neighboring, sp$^2$) & -6.2741\\
CH/CH (second neighboring, sp$^2$) & -1.7077\\ 
CC/CH (first neighboring, sp$^3$)  & 3.3872 \\
CH/CH (first neighboring, sp$^3$)  & 5.3316 \\
CC/CC (first neighboring, sp$^3$)  & 0 \\ 
\hline
bond/angle &  $k_{b\theta}$ (kcal/mol$\cdot$\AA$\cdot$rad) \\
\hline
C--C/C--C--C (sp$^2$ atoms) & 28.8708 \\
C--C/C--C--H (sp$^2$ atoms) & 20.0033 \\
C--H/C--C--H (sp$^2$ atoms) & 24.2183 \\  
C--C/C--C--C (sp$^3$ atoms) & 8.016 \\
C--C/C--C--H (sp$^3$ atoms) & 20.754 \\
C--H/C--C--H (sp$^3$ atoms) & 11.421 \\
C--H/H--C--H (sp$^3$ atoms) & 18.103 \\
\hline
angle/angle &  $k_{\theta\theta'}$ (kcal/mol$\cdot$rad$^2$) \\
\hline
C--C--C/C--C--C (sp$^2$ atoms) & 0 \\
C--C--C/C--C--H (sp$^2$ atoms) & 0 \\
H--C--C/C--C--H (sp$^2$ atoms) & 0 \\ 
C--C--C/C--C--C (sp$^3$ atoms) & -0.1729 \\
C--C--C/C--C--H (sp$^3$ atoms) & -1.3199 \\
H--C--C/C--C--H (sp$^3$ atoms) & -0.4825 \\ 
\hline
angle/angle/torsional & $k_{\theta\theta'\phi}$ (kcal/mol$\cdot$rad$^2$)\\
\hline
C--C--C/C--C--C/C--C--C--C (sp$^2$) &  0 \\
C--C--C/C--C--H/C--C--C--H (sp$^2$) & -4.8141 \\
H--C--C/C--C--H/H--C--C--H (sp$^2$) & ~0.3598 \\
C--C--C/C--C--C/C--C--C--C (sp$^3$) & -22.045 \\
C--C--C/C--C--H/C--C--C--H (sp$^3$) & -16.164 \\
H--C--C/C--C--H/H--C--C--H (sp$^3$) & -12.564 \\ 
\hline
\hline
\end{tabular}
}
\end{table}
\begin{table*}[tbh]
\centering\noindent
\caption{
Values of parameters for cross-coupling bond/torsion potential
$V(b,\phi)=(b-b_0)[k_{b\phi,1}\cos\phi+k_{b\phi,2}\cos2\phi+k_{b\phi,3}\cos3\phi]$.
and cross-coupling angle/torsion potential
$V(\theta,\phi)=(\theta-\theta_0)[k_{\theta\phi,1}\cos\phi+k_{\theta\phi,2}\cos2\phi+k_{\theta\phi,3}\cos3\phi]$.
}
\label{tab3}
{
\begin{tabular}{lccc}
\hline
\hline
Bond/torsion & ~~$k_{b\phi,1}$ (kcal/mol$\cdot$\AA)~~ & ~~$k_{b\phi,2}$ (kcal/mol$\cdot$\AA)~~ 
& ~~$k_{b\phi,3}$ (kcal/mol$\cdot$\AA) \\
\hline
C--C/C--C--C--C (sp$^2$ atoms, central bond) & 27.5989 & -2.3120 & 0 \\
C--C/C--C--C--H (sp$^2$ atoms, central bond) & 0       & -1.1521 & 0 \\
C--H/H--C--C--H (sp$^2$ atoms, central bond) & 0      & 4.8228 & 0 \\
C--C/C--C--C--C (sp$^2$ atoms, terminal bond) & -0.1185 & 6.3204 & 0 \\
C--C/C--C--C--H (sp$^2$ atoms, terminal bond) &  0      & -6.8958 & 0 \\
C--H/C--C--C--H (sp$^2$ atoms, terminal bond) &  0      & -0.4669 & 0 \\
C--H/H--C--C--H (sp$^2$ atoms, terminal bond) &  0      & -0.6890 & 0 \\ 
C--C/C--C--C--C (sp$^3$ atoms, central bond) & -17.787 & -7.1877 & 0 \\
C--C/C--C--C--H (sp$^3$ atoms, central bond) & -14.879 & -3.6581 & -0.3138 \\
C--C/H--C--C--H (sp$^3$ atoms, central bond) & -14.261 & -0.5322 & -0.4864 \\ 
C--C/C--C--C--C (sp$^3$ atoms, terminal bond) & -0.0732 & 0      & 0 \\
C--C/C--C--C--H (sp$^3$ atoms, terminal bond) &  0.2486 & 0.2422 & -0.0925 \\
C--H/C--C--C--H (sp$^3$ atoms, terminal bond) &  0.0814 & 0.0591 & 0.2219 \\
C--H/H--C--C--H (sp$^3$ atoms, terminal bond) & 0.2130  & 0.3120 & 0.0777 \\
\hline
angle/torsion & ~~$k_{\theta\phi,1}$ (kcal/mol$\cdot$rad)~~ & 
~~$k_{\theta\phi,2}$ (kcal/mol$\cdot$rad)~~ 
& ~~$k_{\theta\phi,3}$ (kcal/mol$\cdot$rad) \\
\hline
C--C--C/C--C--C--C (sp$^2$ atoms) & 1.9767 & 1.0239 & 0 \\
C--C--C/C--C--C--H (sp$^2$ atoms) & 0 & 2.5014 & 0 \\
C--C--H/C--C--C--H (sp$^2$ atoms) & 0 & 2.7147 & 0 \\
H--C--C/H--C--C--H (sp$^2$ atoms) & 0 & 2.4501 & 0 \\ 
C--C--C/C--C--C--C (sp$^3$ atoms) & 0.3886 & -0.3139 & 0.1389 \\
C--C--C/C--C--C--H (sp$^3$ atoms) & -0.2454 & 0 & -0.1136 \\
C--C--H/C--C--C--H (sp$^3$ atoms) &  0.3113 & 0.4516 & -0.1988 \\
H--C--C/H--C--C--H (sp$^3$ atoms) & -0.8085 & 0.5569 & -0.2466 \\ 
\hline
\hline
\end{tabular}
}
\end{table*}
\begin{table}[tbh]
\caption{
Values of parameters for LJ-9-6 potential
$V(r)=\epsilon[2(r_0/r)^9-3(r_0/r)^6]$ and
charge valence bond increments $q_1$, $q_2$}
\label{tab4}
\centering\noindent
{
\begin{tabular}{lcc}
\hline
\hline
atom atom & ~~$\epsilon$ (kcal/mol)~~ & ~~$r_0$ (\AA) \\
\hline
 C C (sp$^2$ atoms)   & 0.0680  & 3.9150 \\
 C H (sp$^2$ atoms)   & 0.0271  & 3.5741 \\
 H H (sp$^2$ atoms)   & 0.0230  & 2.878 \\ 
 C C (sp$^3$ atoms)   & 0.0400 & 3.854 \\
 H H (sp$^3$ atoms)   & 0.0230 & 2.878 \\
 F F (sp$^3$ atoms)   & 0.0598 & 3.200 \\
 C H (sp$^3$ atoms)   & 0.0215 & 3.526 \\
 C F (sp$^3$ atoms)   & 0.0422 & 3.600 \\
\hline
atom atom & ~~$q_1$ (e)~~ & ~~$q_2$ (e) \\
\hline
C  C (sp$^2$ atoms)   & 0       &  0  \\
C  H (sp$^2$ atoms)   & -0.1268 & +0.1268  \\
C  C (sp$^3$ atoms)   & 0       &  0  \\
C  H (sp$^3$ atoms)   & -0.053  & +0.053 \\
C  F (sp$^3$ atoms)   & +0.25   & -0.25\\ 
\hline
\hline 
\end{tabular} 
}
\end{table}

Taking into consideration the noncovalent interactions of atoms only from neighboring unit cells, 
the Hamiltonian of a nanoribbon can be written in the following form
\begin{equation}
H=\sum_{n=-\infty}^{+\infty}\left[\frac12({\bf M}\dot{\bf u}_n,\dot{\bf u}_n)
+P({\bf u}_{n-1},{\bf u}_n,{\bf u}_{n+1})\right], \label{f2}
\end{equation}
where $n$ -- number of the unit cell, ${\bf u}_n$ -- $3N_e$-dimensional vector defining coordinates
of atoms in the unit cell, ${\bf M}$ -- the diagonal matrix of cell atomic masses,\linebreak 
$P({\bf u}_{n-1},{\bf u}_n,{\bf u}_{n+1})$ -- of the cell atom's interaction with each other 
and with atoms from neighboring cells.

In the ground state each nanoribbon unit cell is obtained from the previous one by shifting by the step
$a$: ${\bf u}_n^0={\bf u}^0+an{\bf e}_x$, where the unit vector
${\bf e}_x=\{(1,0,0)_k\}_{k=1}^{N_e}$ (nanoribbon lies along the $x$ axes).
To find the ground state (the period $a$ and the coordinate vector ${\bf u}^0$) 
we should solve the minimum problem:
\begin{equation}
P({\bf u}^0-a{\bf e}_x,{\bf u}^0,{\bf u}^0+a{\bf e}_x)\rightarrow\min: {\bf u}^0, a.
\label{f3}
\end{equation}     
The problem (\ref{f3}) is solved by the conjugate gradient method. The numerical solution shows 
that the ground state calculated by the COMPASS force field agrees well with the experimental 
values (the equilibrium values of bond lengths, valence angles and torsional angles are the same). 
Let us verify the coincidence of the of the nanoribbon frequency spectrum with the experimental 
data. To accomplish this, we find the dispersal curve of the nanoribbon.

\section{Dispersion equation}

Let us introduce a $3N_e$-dimensional vector
$$
{\bf v}_n=\{{\bf u}_{(n,k)}-{\bf u}_{(n,k)}^0\}_{k=1}^{N_e}
$$
describing the displacement of atoms in the $n$-th cell from
their equilibrium positions. Then the Hamiltonian (\ref{f2}) of a
nanoribbon can be written in the following form:
\begin{equation}
H=\sum_{n=-\infty}^{+\infty}\left[\frac12({\bf M}\dot{\bf v}_n,\dot{\bf v}_n)
+U({\bf v}_{n-1},{\bf v}_n,{\bf v}_{n+1})\right], \label{f4}
\end{equation}
where interaction potential 
$$
U({\bf v}_{1},{\bf v}_2,{\bf v}_{3})=
P({\bf u}^0-a{\bf e}_x+{\bf v}_1,{\bf u}^0+{\bf v}_2,{\bf u}^0+a{\bf e}_x+{\bf v}_3).
$$

The Hamiltonian (\ref{f4}) corresponds to the equations of motion
\begin{eqnarray}
-{\bf M}\ddot{\bf v}_n&=&U_{{\bf v}_1}({\bf v}_{n},{\bf v}_{n+1},{\bf v}_{n+2})
+U_{{\bf v}_2}({\bf v}_{n-1},{\bf v}_{n},{\bf v}_{n+1})\nonumber\\
&+&U_{{\bf v}_3}({\bf v}_{n-2},{\bf v}_{n-1},{\bf v}_{n}),\label{f5}
\end{eqnarray}
were vector $U_{{\bf v}_i}=\partial U({\bf v}_1,{\bf v}_2,{\bf v}_3)/\partial{\bf v}_i$, $i=1,2,3$.

For small displacements, system (\ref{f5}) can be written as a system
of linear equations
\begin{equation}
-{\bf M}\ddot{\bf v}_n={\bf B}_1{\bf v}_{n}+{\bf B}_2{\bf v}_{n+1}+{\bf B}_2^*{\bf v}_{n-1}
+{\bf B}_3{\bf v}_{n+2}+{\bf B}_3^*{\bf v}_{n-2}, \label{f6}
\end{equation}
where the matrices 
${\bf B}_1=U_{{\bf v}_1,{\bf v}_1}+U_{{\bf v}_2,{\bf v}_2}+U_{{\bf v}_3,{\bf v}_3}$,
${\bf B}_2=U_{{\bf v}_1,{\bf v}_2}+U_{{\bf v}_2,{\bf v}_3}$,
${\bf B}_3=U_{{\bf v}_1,{\bf v}_3}$ and the matrices of partial derivatives
are
$$
U_{{\bf v}_i,{\bf v}_j}=\frac{\partial^2U}{\partial{\bf v}_i\partial{\bf v}_j}
({\bf 0},{\bf 0},{\bf 0}),~~i,j=1,2,3.
$$

The solution of the systems of linear equations (\ref{f6}) can
be written in the standard form of the wave
\begin{equation}
{\bf v}_n=A{\bf w}\exp(iqn-i\omega t), \label{f7}
\end{equation}
where $A$ -- amplitude, ${\bf w}$ -- eigenvector,  $\omega$ is the phonon frequency with the dimensionless
wave number $q\in[0,\pi]$. Substituting Eq. (\ref{f7}) into Eq. (\ref{f6}), we
obtain the eigenvalue problem
\begin{equation}
\omega^2{\bf M}{\bf w}={\bf C}(q){\bf w}, \label{f8}
\end{equation}
where Hermitian matrix 
$$
{\bf C}(q)={\bf B}_1+{\bf B}_2e^{iq}+{\bf B}_2^*e^{-iq}+{\bf B}_3e^{2iq}+{\bf B}_3^*e^{-2iq}.
$$
Using the substitution ${\bf w}={\bf M}^{-1/2}{\bf e}$, problem (\ref{f8}) can be
rewritten in the form
\begin{equation}
\omega^2{\bf e}={\bf M}^{-1/2}{\bf C}(q){\bf M}^{-1/2}{\bf e},
\label{f9}
\end{equation}
where ${\bf e}$ is the normalized eigenvector, $({\bf e},{\bf e})=1$.

Thus, for obtaining the dispersion curves $\omega_j(q)$, it is necessary
to find the eigenvectors of the Hermitian matrix (\ref{f9})
of a size $3N_e\times 3N_e$ for each fixed wave number $0\le q\le\pi$.
As a result, we obtain $3N_e$ branches of the dispersion curve
$\{\omega_j(q)\}_{j=1}^{3N_e}$.
\begin{figure}[t]
\begin{center}
\includegraphics[angle=0, width=1.0\linewidth]{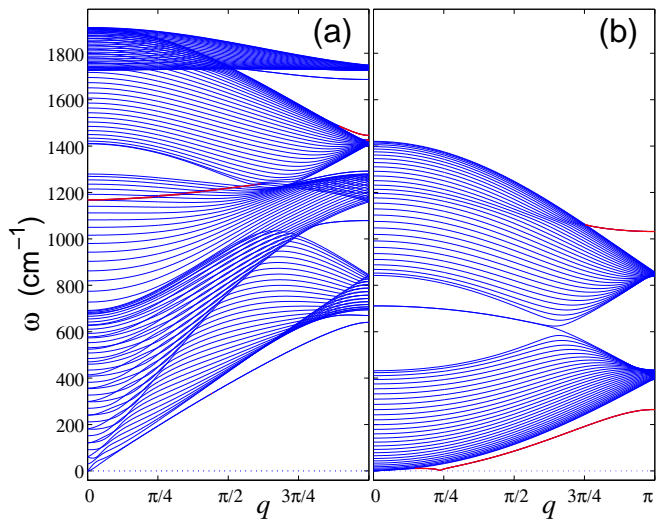}
\end{center}
\caption{
Structure of $3N_e$ dispersion curves for zigzag graphene nanoribbon (C$_{64}$H$_2$)$_\infty$
for (a) in-plane and (b) out-of-plain vibrations.
The thick red curves correspond to oscillations localized at the nanoribbon edges.   
}
\label{fig3}
\end{figure}

\section{Graphene frequency spectrum}
Let us consider a wide nanoribbon (C$_{64}$H$_2$)$_\infty$ (the nano\-ribbon width
$D=68.79$\AA, the period $a=2.412$\AA, the number of atoms in unit cell $N_e=66$)
to find the frequency spectrum of graphene sheet. Figure \ref{fig3} shows the $3N_e$ 
dispersion curves of the nanoribbon. The plane structure of the nanoribbon allows to divide 
its vibrations into two classes: in-plane vibrations, when the atoms are always stayed 
in the plane of the nanoribbon and out-of-plane vibrations when the atoms are shifted 
orthogonal to the plane. Two third of the branches corresponds to the atom vibrations 
in the $xy$ plane of the nanoribbon (in-plane vibrations), whereas only one third corresponds 
to the vibrations orthogonal to the plane (out-of-plane vibrations), when the atoms are 
shifted along the axes $z$.

The nanoribbon frequency spectrum consists of two intervals $[0,1911]$ and 
$[3086.6,3099.2]$~cm$^{-1}$ (the second high-frequency interval corresponds to the edge 
vibrations of the valence bonds C--H). Discarding frequencies of the edge vibrations from 
the Figure \ref{fig3}, we can conclude that graphene sheet spectrum consists of the one 
frequency interval $[0, \omega_m]$ with the maximum frequency $\omega_m=1911$~cm$^{-1}$. 
Spectrum of out-of-plane vibrations also consists of the one frequency interval 
$[0,\omega_o]$ with the maximum frequency $\omega_o=1420$~cm$^{-1}$.

The frequency spectrum of graphene sheet has been studied theoretically and experimentally 
in \cite{pm57,pm58,pm59,pm60}. The maximum frequency of in-plane vibrations is $\omega_m=1600$~cm$^{-1}$,
the maximum frequency of out-of-plane vibrations is $\omega_o=868$~cm$^{-1}$. Thus, 
the graphene frequency spectrum calculated by the COMPASS force field does not coincide with 
experimental valuations. The frequency spectrum of in-plane vibrations is 1.2 times more and 
the frequency spectrum of out-of-plane vibrations is 1.6 times more than experimental valuations, 
i.e. the bending stiffness of the nanoribbon is 2.5 times more.
\begin{figure}[t]
\begin{center}
\includegraphics[angle=0, width=1.0\linewidth]{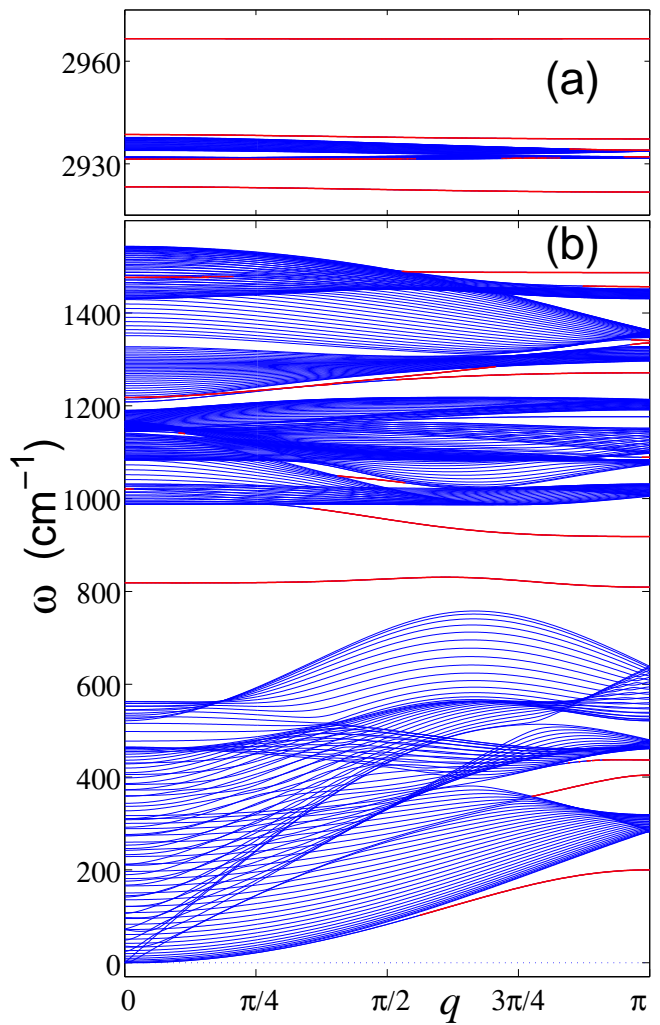}
\end{center}
\caption{
Structure of $3N_e$ dispersion curves for zigzag graphane nanoribbon (C$_{64}$H$_{66}$)$_\infty$.
The thick red curves correspond to oscillations localized at the nanoribbon edges.   
}
\label{fig4}
\end{figure}

\section{Graphane and fluorographene frequency spectrum}
Let us consider a wide nanoribbon
(C$_{64}$H$_{66}$)$_\infty$ (nanoribbon width $D=68.4$\AA, period $a=2.522$\AA, 
number atoms in unit cell $N_e=130$) to find the frequency spectrum of graphane sheet.
Figure \ref{fig4} shows the dispersion curves of the nanoribbon.
As can be seen from the figure, the nanoribbon frequency spectrum consists of three intervals: 
low-frequency interval $[0,757.8]$, middle-frequency interval $[987.5,1543.6]$
and narrow\linebreak high-frequency interval $[2921.8,2966.6]$~cm$^{-1}$.
On average, carbon atoms account for 79.5\% of the vibration energy in the first frequency interval,
for 48.1\% in the second frequency interval and only for 8.4\% in the third. 
It enables us to say that the low-frequency spectrum corresponds to the vibrations, 
in which the valence bond C--H and the valence angles C--C--H remain nearly unchanged, 
the middle interval corresponds to the vibrations, in which the valence corners C--C--H 
begin to take part in the vibration, and the third interval corresponds to the vibrations 
of the hard valence bonds C--H (the number of such modes always coincide with the number 
of hydrogen atoms).

Since we have considered as sufficiently broad nano\-ribbon, the analysis of its dispersal 
curves suggests that the frequency spectrum of linear vibrations of the endless 
graphene sheet (CH)$_\infty$  consists of three continuous intervals 
$[0,756]$, $[987,1544]$ and $[2922, 2967]$~cm$^{-1}$.
Thus, the graphene sheet has two gaps in the frequency spectrum: narrow low-frequency $[756,987]$ 
and wide\linebreak 
high-frequency $[1544,2922]$~cm$^{-1}$. Such structure of the frequency spectrum is 
in good agreement with the results of first-principles calculations. 
Quantum calculations suggest the high-frequency spectrum intervals\linebreak 
$[0,806]$, $[950,1350]$ and $[2733,2783]$~cm$^{-1}$ \cite{pm61}.
\begin{figure}[t]
\begin{center}
\includegraphics[angle=0, width=1.0\linewidth]{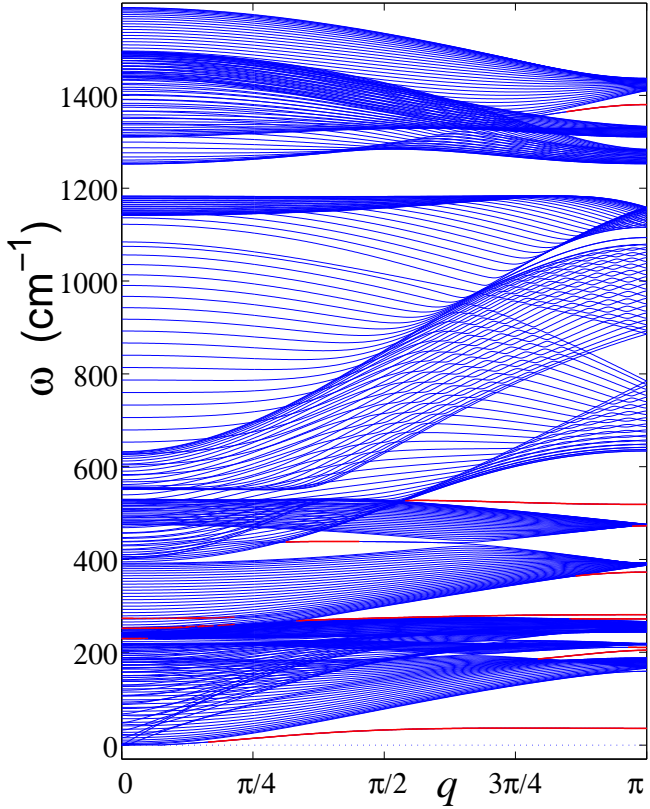}
\end{center}
\caption{
Structure of $3N_e$ dispersion curves for zigzag graphene fluoride nanoribbon (C$_{64}$F$_{66}$)$_\infty$.
The thick red curves correspond to oscillations localized at the nanoribbon edges.   
}
\label{fig5}
\end{figure}

If the hydrogen atoms H change to the heavier deuterium atoms D, the three-zoned structure 
of the frequency spectrum will remain, but all the frequencies will shift down. 
The frequency spectrum of graphene sheet (CD)$_\infty$ also consists of three intervals 
$[0,620]$, $[813,1503]$, $[2130,2200]$~cm$^{-1}$ (the high-frequency interval most strongly shifts down).
The stronger increase in the weight of the connected atoms, the change of hydrogen atoms 
for fluorine atoms, causes the stronger frequency shift and the closure of the gaps. 

Figure \ref{fig5} shows the dispersion curves for fluoro\-gra\-phene nanoribbon (C$_{64}$F$_{66}$)$_\infty$.
The figure clearly shows that the nanoribbon frequency spectrum of linear vibrations consists 
of two intervals: $[0,1183.6]$ and\linebreak $[1252.0,1589.3]$~cm$^{-1}$.
The analysis of the dispersion curves suggests that the frequency spectrum of linear vibrations 
of the endless fluorographene sheet (CF)$_\infty$ consists of two continuous intervals 
$[0,1184]$ and\linebreak $[1252,1589]$~cm$^{-1}$ with the narrow gap between them. Such structure of the 
frequency spectrum is also in good agreement with the results of quantum chemical calculations \cite{pm61}.
Quantum calculations suggest the frequency spectrum intervals
$[0,1039]$ and\linebreak $[1106,1312]$~cm$^{-1}$.
\begin{figure}[t]
\begin{center}
\includegraphics[angle=0, width=1.0\linewidth]{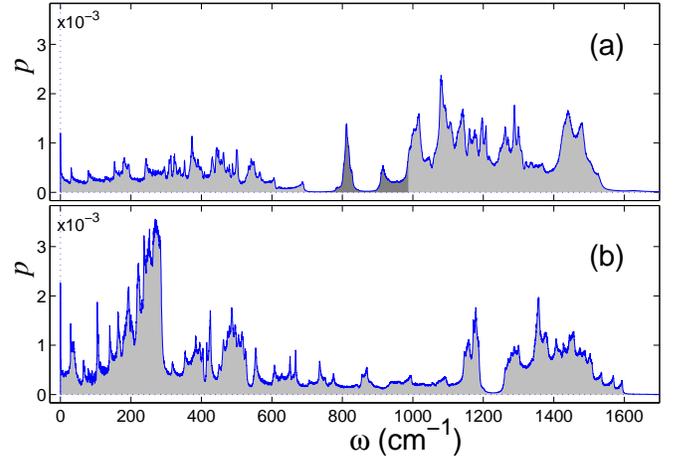}
\end{center}
\caption{
Density of states $p(\omega)$ of atom vibrations in (a) graphane nanoribbon
(C$_{12}$H$_{14}$)$_{99}$C$_{10}$H$_{22}$ 
(darker color corresponds to frequency region of the gap edge modes)
and (b) graphene fluoride nanoribbon
(C$_{12}$F$_{14}$)$_{99}$C$_{10}$F$_{22}$ for temperature $T=300$K.
The density of states is normalized by the condition
$\int_0^\infty p(\omega)d\omega=1$.     
}
\label{fig6}
\end{figure}

The structure of the frequency spectrum can be obtained also from the analysis 
of the frequency spectrum density of atomic thermal vibrations of finite length nanoribbon. 
Let us consider the graphane nanoribbon (C$_{12}$H$_{14}$)$_{99}$C$_{10}$H$_{22}$ with size 
$25.09\times 1.17$~nm$^2$, consisted of $N=100$ unit cells. 

The dynamics of the thermalized nanoribbon is described by the system of Langevin equations
\begin{eqnarray}
M_{n,k}\ddot{\bf u}_{n,k}=-\frac{\partial H}{\partial{\bf u}_{n,k}}-\Gamma M_{n,k}+\Xi_{n,k},
\label{f10} \\
n=1,2,...,N,~~k=1,2,...,N_e, \nonumber
\end{eqnarray}
where $H$ -- Hamiltonian of the nanoribbon, created by the COMPASS force field,
${\bf u}_{n,k}=(u_{n,k,1}, u_{n,k,2}, u_{n,k,3})$ -- 3D coordinate vector of atom
$(n,k)$, $M_{n,k}$ -- mass of this
atom, $\Gamma=1/t_r$ -- damping coefficient 
(relaxation time $t_r=0.4$ ps). 
Normally distributed random forces
$\Xi_{n,k}=(\xi_{n,k,1},\xi_{n,k,2},\xi_{n,k,3})$ normalized by conditions
$$
\langle\xi_{n,k,i}(t_1)\xi_{m,l,j}(t_2)\rangle=2M_{n,k}\Gamma k_BT\delta_{nm}\delta_{kl}\delta_{ij}\delta(t_1-t_2),
$$
where $k_B$ -- Boltzmann's constant, $T$ -- thermostat temperature.

The system of equations of motion (\ref{f10}) was integrated numerically during the time $t= 10$~ps
(the stationary state of the plane nanoribbon serves as a starting point). During this time, 
the nanoribbon and the thermostat came to balance. Then its interaction with the thermostat 
was switched off and the dynamics of isolated thermalized nanoribbon was considered. 
The density of the frequency spectrum of atomic vibrations $p(\omega)$ was calculated via 
fast Fourier transform.

The density of the frequency spectrum of atomic vibrations of the graphane and fluorographene 
nanoribbons at $T=300$~K is presented in Fig.~\ref{fig6}. The figure clearly shows that 
the profile of density $p(\omega)$ coincides well with the forms of the dispersal curves. 
The low-frequency gap is clearly visible for the graphane nanoribbon, and the narrow gap 
in the frequency spectrum is clearly visible for the fluorographene nanoribbon.

\section{Corrections of the COMPASS force field for graphene sheet}
The simulation shows that the COMPASS force field does not describe well the frequency spectrum 
of gra\-phene sheet. The frequency spectrum of out-of-plane vibrations differs strongly from 
the experimental values. 
\begin{figure}[t]
\begin{center}
\includegraphics[angle=0, width=1.0\linewidth]{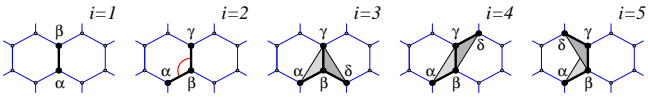}
\end{center}
\caption{
Configurations containing up to $i$-th nearest-neighbor interactions in graphene sheet.
}
\label{fig7}
\end{figure}

The potential energy of the graphene sheet depends on variations 
in bond length, bond angles, and dihedral angles between the planes formed by three neighboring 
carbon atoms and it can be written in the form
\begin{equation}
E=\sum_{\Omega_1}V_1+\sum_{\Omega_2}V_2+\sum_{\Omega_3}V_3+\sum_{\Omega_4}V_4+\sum_{\Omega_5}V_5,
\label{f11}
\end{equation}
where $\Omega_i$, with $i=1,2,3,4,5$, are the sets of configurations including up to 
nearest-neighbor interactions. Owing to a large redundancy, the sets only need to contain 
configurations of the atoms shown in Fig. \ref{fig7}, including their rotated and
mirrored versions.
The potential\linebreak 
$V_1({\bf u}_\alpha,{\bf u}_\beta)$ describes the deformation energy due to a direct 
interaction between pairs of atoms with the indexes $\alpha$ and $\beta$.
The potential $V_2({\bf u}_\alpha,{\bf u}_\beta,{\bf u}_\gamma)$
describes the deformation energy of the angle between the valent bonds ${\bf u}_\alpha{\bf u}_\beta$
and ${\bf u}_\beta{\bf u}_\gamma$.
\begin{figure}[t]
\begin{center}
\includegraphics[angle=0, width=1.0\linewidth]{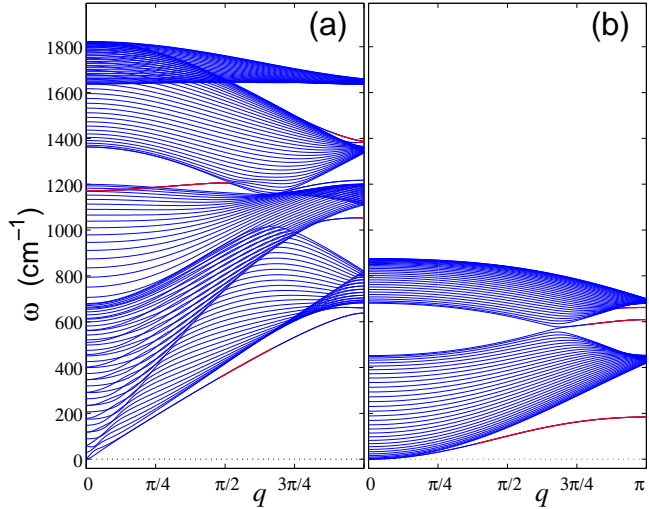}
\end{center}
\caption{
Structure of $3N_e$ dispersion curves for zigzag graphene nanoribbon (C$_{64}$H$_2$)$_\infty$
for (a) in-plane and (b) out-of-plain vibrations.
The thick red curves correspond to oscillations localized at the nanoribbon edges.   
The force field COMPASS with correction of parameters (\ref{f12}) was used.
}
\label{fig8}
\end{figure}

Potentials $V_i({\bf u}_\alpha,{\bf u}_\beta,{\bf u}_\gamma,{\bf u}_\delta)$, $i = 3,4,5$, 
describes the deformation energy associated with a change of the effective angle between 
the planes ${\bf u}_\alpha{\bf u}_\beta{\bf u}_\gamma$ and ${\bf u}_\beta{\bf u}_\gamma{\bf u}_\alpha$
as shown in Fig.~\ref{fig7}.

In the COMPASS force field the first potential
$V_1=k_{b,2}(b-b_0)^2+k_{b,3}(b-b_0)^3+k_{b,4}(b-b_0)^4$
describes the deformations of valence bonds, the second potential
$V_2=k_{\theta,2}(\theta-\theta_0)^2+k_{\theta,3}(\theta-\theta_0)^3+k_{\theta,4}(\theta-\theta_0)^4$
-- the deformations of valence angles, the third potential $V_3(\chi)=k_{\chi}\chi^2$ -- 
out-of-plane deformations.
The potentials of dihedral angles $V_4$ and $V_5$ are described by a single potential of
the torsional angle
$V_t(\phi)=k_{\phi,1}(1-\cos\phi)+k_{\phi,2}(1-\cos2\phi)+k_{\phi,3}(1-\cos3\phi)$.

The specific value of the parameter $\kappa_{\chi}=5.7537$ kcal/mol can be
found from the frequency spectrum of small-amplitude oscillations of a
sheet of graphite \cite{sav1}. 
For the potentials of dihedral angles
$V_4=c_4(\phi-\pi)^2$ and $V_5=c_5\phi^2$
according to the results of Ref. \cite{sav2} 
coefficient $c_4$ is close to $k_\chi$, whereas $c_5\ll c_4$
($|c_5/c_4|<1/20$). So for the second derivative of the torsion potential we have 
\begin{eqnarray}
V_t''(0)&=&k_{\phi,1}+4k_{\phi,2}=V_5''(0)=0, \nonumber\\
V_t''(\pi)&=&-k_{\phi,1}+4k_{\phi,2}=V_4''(0)=2c_4=2k_\chi, \nonumber
\end{eqnarray}
therefore coefficients 
$k_{\phi,2}=k_\chi/4=1.4384$ kcal/mol, $k_{\phi,1}=-4k_{\phi,2}=-5.7537$~kcal/mol. 

Now let us consider the COMPASS force field with the new parameters values
\begin{equation}
k_{\chi}=-k_{\phi,1}=5.7537,~~k_{\phi,2}=1.4384\mbox{kcal/mol}.
\label{f12}
\end{equation}
Since all the parameters for the off-diagonal interactions bond/torsion, angle/torsion 
and angle/angle/torsion have become undefined after such a modification, remove them away 
from the force field. Then we find the frequency value of the graphene sheet via 
that modification of the force field.

Figure 8 shows the dispersal curves of the graphene nanoribbon (C$_{64}$H$_2$)$_\infty$ 
calculated via the modified COMPASS force field.
From figure, we can conclude that now the spectrum of the graphene sheet consists 
of the frequency interval of in-plane vibrations $[0,1823]$ and the frequency interval 
of out-of-plane vibrations $[0,876]$~cm$^{-1}$, that is in good agreement with 
the results described in \cite{pm57,pm58,pm59,pm60}.
Thus, modification (\ref{f12}) of the COMPASS force field allows to take into account  
the flexural mobility of graphene sheet (the initial set of parameters led to an 
over estimation of the bending stiffness)

\section{Conclusion}

The study shows that the COMPASS force field allows to describe with good accuracy 
the frequency spectrum of atomic vibrations of graphane and fluorographene sheets, 
whose honeycomb hexagonal lattice is formed by sp$^3$ valence bands. 
On the other hand, the force field doesn't describe very well the frequency spectrum 
of graphene sheet, whose planar hexagonal lattice is formed by sp$^2$ valence bands. 
In that case the frequency spectrum of out-of-plane vibrations differs greatly 
from the experimental data (bending stiffness of a graphene sheet is strongly overestimated). 
The correction (\ref{f12}) of parameters of out-of-plane and torsional potentials 
of the force field was made, and that allows to achieve the coincidence of vibration 
frequency with experimental data. After such corrections the COMPASS force field 
can be used to describe the dynamics of flat graphene sheets and carbon nanotubes. 

\section*{Acknowledgements}

This work is supported by the Russian Science Foundation under grant 16-409 13-10302. 
The research was carried out using supercomputers at Joint Supercomputer Center 
of 410 the Russian Academy of Sciences (JSCC RAS).


\begin{references}
\bibitem{pm1}
Novoselov KS, Geim AK, Morozov SV, Jiang D, Zhang Y, Dubonos SV, Grigorieva IV,
 Firsov AA (2004) Electric field effect in atomically thin carbon films. 
Science 306:666-669. https://doi.org/10.1126/science.1102896
\bibitem{pm2}
Mohan VB, Lau K, Hui D, Bhattacharyya D (2018)  
Graphene-based materials and their composites: A review on production, applications and product limitations.
Composites B 142:200-220. https://doi.org/10.1016/j.compositesb.2018.01.013
\bibitem{pm3}
Akinwande D, Brennan CJ, Bunch JS et al (2017) A review on mechanics and mechanical properties of 2D materials-Graphene and beyond. 
Extreme Mechanics Letters 13:42-77. http://dx.doi.org/10.1016/j.eml.2017.01.008
\bibitem{pm4}
Harik V (2018) Mechanics of Carbon Nanotubes. Fundamentals, Modelling and Safety. Academic Press. 
https://doi.org/10.1016/C2016-0-00799-4
\bibitem{pm5}
Feng W, Long P, Feng Y, Li Y (2016)
Two-Dimensional Fluorinated Graphene: Synthesis, Structures, Properties and Applications.
Adv. Sci. 3, 1500413. https://doi.org/10.1002/advs.201500413
\bibitem{pm6}
Chen T, Cheung R (2016) Mechanical properties of graphene. 
In: Aliofkhazraei M, Ali N, Milne WI, Ozkan CS, Mitura S, Gervasoni JL (eds.) 
Graphene Science Handbook: Mechanical and Chemical Properties. CRC Press, London. pp. 3-15 
\bibitem{pm7}
Bhimanapati GR, Lin Z, Meunieret V et al (2015) 
Recent advances in two-dimensional materials beyond graphene. 
ACS Nano 9:11509-115399. https://doi.org/10.1021/acsnano.5b05556
\bibitem{pm8}
Cao G (2014) 
Atomistic Studies of Mechanical Properties of Graphene. Review. 
Polymers 6:2404-2432. https://doi.org/10.3390/polym6092404
\bibitem{pm9}
Li X, Tao L, Chen Z, Fang H, Li X, Wang X, Xu J-B, Zhu H (2017) 
Graphene and related two-dimensional materials: Structure-property relationships 
for electronics and optoelectronics. 
Appl. Phys. Rev. 4, 021306 (2017). http://dx.doi.org/10.1063/1.4983646
\bibitem{pm10}
Geim AK, Novoselov KS (2007) The rise of graphene. Nat. Mater. 6:183-191. https://doi.org/10.1038/nmat1849
\bibitem{pm11}
Brar VW, Sherrott MC, Jariwala D (2018)  
Emerging photonic architectures in two-dimensional opto-electronics. 
Chem. Soc. Rev. 47:6824-6844. https://doi.org/10.1039/c8cs00206a
\bibitem{pm12}
Khan ME, Khan MM, Cho MH (2018) 
Recent Progress of Metal-Graphene Nanostructures in Photocatalysis. 
Nanoscale 10: 9427-9440. https://doi.org/10.1039/C8NR03500H
\bibitem{pm13}
Ferrari AC, Bonaccorso F, Fal'ko V, et al (2015)  
Science and technology roadmap for graphene, related two-dimensional crystals, and hybrid systems.
Nanoscale 7:4598-5062. https://doi.org/10.1039/c4nr01600a
\bibitem{pm14}
Xiang Q, Cheng B, Yu J (2015)  
Graphene-Based Photocatalysts for Solar-Fuel Generation. 
Angew. Chem., Int. Ed. 54:11350-11366. https://doi.org/10.1002/anie.201411096 
\bibitem{pm15}
Ryu J, Lee  E, Lee  K, Jang  J (2015) 
A Graphene Quantum Dots Based Fluorescent Sensor for Anthrax Biomarker Detection and Its Size Dependence.
J. Mater. Chem. B 3:4865-4870 https://doi.org/10.1039/C5TB00585J
\bibitem{pm16}
Soldano C, Mahmood A, Dujardin E (2010) 
Production, properties and potential of graphene. 
Carbon  48:2127-2150. https://doi.org/10.1016/j.carbon.2010.01.058
\bibitem{pm17}
Li X, Zhi L (2018) 
Graphene hybridization for energy storage applications. 
Chem. Soc. Rev. 47:3189-3216. https://doi.org/10.1039/C7CS00871F
\bibitem{pm18}
Siahlo AI, Poklonski NA, Lebedev AV, Lebedeva IV, Popov AM, Vyrko SA,
Knizhnik AA, Lozovik YE (2018)
Structure and energetics of carbon, hexagonal boron nitride, and carbon/hexagonal
boron nitride single-layer and bilayer nanoscrolls.
Phys. Rev. Mater. 2, 036001. https://doi.org/10.1103/PhysRevMaterials.2.036001
\bibitem{pm19}
Azadmanjiri J, Srivastava VK, Kumar P, Nikzad M, Wang J, Yu A (2018) 
Two- and three-dimensional graphene-based hybrid composites 
for advanced energy storage and conversion devices. 
J. Mater. Chem. A 6, 702-734. https://doi.org/10.1039/C7TA08748A
\bibitem{pm20}
Sunnardianto GK, Maruyama I, Kusakabe K (2017) 
Storing-hydrogen processes on graphene activated by atomic-vacancies. 
International Journal of Hydrogen Energy 42:23691-23697. https://doi.org/10.1016/j.ijhydene.2017.01.115
\bibitem{pm21}
Stankovich S, Dikin DA, Dommett GHB, Kohlhaas KM, Zimney KM, Stach EA, 
Piner RD, Nguyen ST, Ruoff RS (2006) 
Graphene-based composite materials. 
Nature 442:282-286. https://doi.org/10.1038/nature04969
\bibitem{pm22}
Tan KH, Sattari S, Donskyi IS, Cuellar-Camacho JL, Cheng C, Schwibbert K, Lippitz A, 
 Unger WES, Gorbushina A, Adeli M, Haag R (2018) 
Functionalized 2D nanomaterials with switchable binding to investigate graphene-bacteria interactions.
Nanoscale 10:9525-9537. https://doi.org/10.1039/c8nr01347k
\bibitem{pm23}
Cheng C, Li S, Thomas A, Kotov NA, Haag R (2017) 
Functional Graphene Nanomaterials Based Architectures: Biointeractions, Fabrications, 
and Emerging Biological Applications. 
Chem. Rev. 117:1826-9537. https://doi.org/10.1039/c8nr01347k
\bibitem{pm24}
Brenner DW (1990) 
Empirical potential for hydrocarbons for use in simulating the chemical vapor 
deposition of diamond films. 
Phys. Rev. B 42:9458-9471. https://doi.org/10.1103/PhysRevB.42.9458 
\bibitem{pm25}
Stuart SJ, Tutein AB, Harrison JA (2000) 
A reactive potential for hydrocarbons with intermolecular interactions. 
J. Chem. Phys. 112:6472. https://doi.org/10.1063/1.481208
\bibitem{pm26}
Brenner DW, Shenderova OA, Harrison JA, Stuart SJ, Ni B, Sinnott SB (2002)
A second-generation reactive empirical bond order (REBO) potential energy expressi2on for hydrocarbons.
J. Phys.: Condens. Matter 14:783-802. https://doi.org/10.1088/0953-8984/14/4/312  
\bibitem{pm27}
Lindsay L, Briodo DA (2010) 
Optimized Tersoff and Brenner empirical potential parameters for lattice dynamics 
and phonon thermal transport in carbon nanotubes and graphene.
Phys. Rev. B 81:205441. https://doi.org/10.1103/PhysRevB.81.205441
\bibitem{pm28}
Senftle TP, Hong S, Islam MM and et al (2016) 
The ReaxFF reactive force-field: Development, applications and future Directions. 
npj Comput. Mater. 2:15011 (2016). https://doi.org/10.1038/npjcompumats.2015.11
\bibitem{pm29}
Los JH, Ghiringhelli LM, Meijer EJ, Fasolino A (2005) 
Improved long-rangereactive bond-order potential for carbon. I. Construction. 
Phys Rev. B 72, 214102. https://doi.org/10.1103/PhysRevB.72.214102
\bibitem{pm30}
Zhou XW, Ward DK, Foster ME (2015) 
An analytical bond-order potential for carbon. 
J Comput Chem. 36:11719-1735. https://doi.org/10.1002/jcc.23949
\bibitem{pm31}
Marks NA (2001)
Generalizing the environment-dependent interaction potential for carbon. 
Phys Rev B. 63, 035401. https://doi.org/10.1103/PhysRevB.63.035401
\bibitem{pm32}
Uddin J, Baskes MI, Srinivasan SG, Cundari TR, Wilson AK (2010)
Modified embedded atom method study of the mechanical properties of carbon nanotube 
reinforced nickel composites. 
Phys Rev B. 81, 104103. https://doi.org/10.1103/PhysRevB.81.104103
\bibitem{pm33}
Mayo SL, Olafson BD, Goddard III WA (1990) 
DREIDING: a generic force field for molecular simulations. 
J. Phys. Chem. 94:8897-8909. https://doi.org/10.1021/j100389a010
\bibitem{pm34}
Tserpes KI, Papanikos P (2014) 
Finite element modeling of the tensile behavior of carbon nanotubes, graphene and 
their composites. In: Tserpes KI, Silvestre N (eds.) 
Springer Series in Materials Science, 
vol. 188: Modeling of Carbon Nanotubes, Graphene and their Composites, 
Springer International Publishing, Champp. 303-329
\bibitem{pm35}
Kalosakas G, Lathiotakis NN, Galiotis C, Papagelis K (2013) 
In-plane force fields and elastic properties of graphene. 
J. APP. Phys. 113, 134307. https://doi.org/10.1063/1.4798384
\bibitem{pm36}
Fthenakis ZG, Kalosakas G, Chatzidakis GD, Galiotis C, Papagelis K, Lathiotakis NN (2017) 
Atomistic potential for graphene and other sp$^2$ carbon systems. 
Phys. Chem. Chem. Phys. 19, 30925-30932. https://doi.org/10.1039/c7cp06362h
\bibitem{pm37}
Chatzidakis GD, Kalosakas G, Fthenakis ZG, Lathiotakis NN (2018) 
A torsional potential for graphene derived from fitting to DFT results. 
Eur. Phys. J. B 91, 11. https://doi.org/10.1140/epjb/e2017-80444-5 
\bibitem{pm38}
Zalizniak VE, Zolotov OA (2017) 
Efficient embedded atom method interatomic potential for graphite and carbon nanostructures. 
Molecular Simulation 43:1480-1484. http://dx.doi.org/10.1080/08927022.2017.1324957
\bibitem{pm39}
Korobeynikov SN, Alyokhin VV, Babichev AV (2018) 
Simulation of mechanical parameters of graphene using the DREIDING force field. 
Acta Mech  229:2343-2378. https://doi.org/10.1007/s00707-018-2115-5
\bibitem{pm40}
Sun H (1998)
COMPASS: An ab initio force-field optimized for condensed-phase applications --
Overview with details on alkane and benzene compounds.
J. Phys. Chem. B 102:7338-7364. https://doi.org/10.1021/jp980939v
\bibitem{pm41}
Sun H, Jin Z, Yang C, Akkermans RL, Robertson SH, Spenley NA, Miller S, Todd SM (2016) 
COMPASS II: extended coverage for polymer and druglike molecule databases. 
J. Mol. Model. 22:47. https://doi.org/10.1007/s00894-016-2909-0
\bibitem{pm42}
Cao GX, Chen X (2007)
The effects of chirality and boundary conditions on the mechanical properties 
of single-walled carbon nanotubes. 
Int. J. Solids Struct. 44:5447-5465. https://doi.org/10.1016/j.ijsolstr.2007.01.005
\bibitem{pm43}
Zhang J, He X, Yang L, Wu G, Sha J, Hou C, Yin C, Pan A, Li Z, Liu Y (2013) 
Effect of Tensile Strain on Thermal Conductivity in Monolayer Graphene Nanoribbons: 
A Molecular Dynamics Study. Sensors 13:9388-9395. https://doi.org/10.3390/s130709388
\bibitem{pm44}
Zhou LX, Wang YG, Cao GX (2013) 
Elastic properties of monolayer graphene with different chiralities. 
J. Phys. Condens. Mater. 25, 125302. https://doi.org/10.1088/0953-8984/25/12/125302
\bibitem{pm45}
Shen X, Lin XY, Yousefi N, Jia JJ, Kim JK (2014) 
Wrinkling in graphene sheets and graphene oxide papers. 
Carbon 66:84-92. https://doi.org/10.1016/j.carbon.2013.08.046
\bibitem{pm46}
Melro LS, Pyrz R, Jensen LR (2016) 
A molecular dynamics study on the interaction between epoxy and functionalized graphene sheets. 
Materials Science and Engineering 139, 012036. https://doi.org/10.1088/1757-899X/139/1/012036
\bibitem{pm47}
Savin AV, Mazo MA (2017) 
Molecular Dynamics Simulation of Two-Sided Chemical Modification of Carbon Nanoribbons on a Solid Substrate.
Dokl. Phys. Chem. 473:37-40. https://doi.org/10.1134/S0012501617030022
\bibitem{pm48}
A.V. Savin (2018)
Edge Vibrations of Graphane Nanoribbons.
Phys. Solid State 60:1046-1053.  https://doi.org/10.1134/S1063783418050281 
\bibitem{pm49}
Savin AV, Sakovich RA, Mazo MA (2018) 
Using spiral chain models for study of nanoscroll structures. 
Phys. Rev. B 97, 165436. https://doi.org/10.1103/PhysRevB.97.165436
\bibitem{pm50}
Sluiter MHF, Kawazoe Y (2003) 
Cluster expansion method for adsorption: Application to hydrogen chemisorption on graphene.
Phys. Rev. B 68, 085410. https://doi.org/10.1103/PhysRevB.68.085410
\bibitem{pm51}
Sofo JO, Chaudhari AS, Barber  GD (2007)
Graphane: A two-dimensional hydrocarbon.
Phys. Rev. B 75, 153401. https://doi.org/10.1103/PhysRevB.75.153401
\bibitem{pm52}
Elias DC, Nair RR, Mohiuddin TMG et al (2009)
Control of Graphene's Properties by Reversible Hydrogenation: Evidence for Graphane.
Science 323:610-613. https://doi.org/10.1126/science.1167130
\bibitem{pm53}
Samarakoon D.K., Wang X-Q (2009)
Chair and Twist-Boat Membranes in Hydrogenated Graphene.
ACS Nano 3:4017-4022. https://doi.org/10.1021/nn901317d
\bibitem{pm54}
Yang YE, Yang Y-R, Yan X-H (2012) 
Universal optical properties of graphane nanoribbons: A first-principles study
Physica E 44:1406-1409. https://doi.org/10.1016/j.physe.2012.03.002
\bibitem{pm55}
Peng Q, Dearden AK, Crean J, Han L, Liu S, Wen X, De S (2014)
New materials graphyne, graphdiyne, graphone, and graphane: review of properties, 
synthesis, and application in nanotechnology. 
Nanotechnol. Sci. App. 7:1-29. https://doi.org/10.2147/NSA.S40324
\bibitem{pm56}
Sahin H, Leenaerts O, Singh SK, Peeters FM (2015) 
Graphane. WIREs Comput Mol Sci 5:255-272. https://doi.org/10.1002/wcms.1216
\bibitem{pm57}
Al-Jishi R, Dresselhaus G (1982)
Lattice-dynamical model for graphite.
Phys. Rev. B 26:4514-4522. https://doi.org/10.1103/PhysRevB.26.4514
\bibitem{pm58}
Aizawa T, Souda R, Otani S, Ishizawa Y, Oshima C (1990)
Bond softening in monolayer graphite formed on transition-metal carbide surfaces.
Phys. Rev. B 42:11469-11478. https://doi.org/10.1103/PhysRevB.42.11469
\bibitem{pm59}
Maultzsch J, Reich S, Thomsen C, Requardt H, Ordejon P (2004)
Phonon Dispersion in Graphite.
Phys. Rev. Lett. 92, 075501. https://doi.org/10.1103/PhysRevLett.92.075501
\bibitem{pm60}
Mohr M, Maultzsch J, Dobardzic E et al (2007)
Phonon dispersion of graphite by inelastic x-ray scattering.
Phys. Rev. B 76, 035439. https://doi.org/10.1103/PhysRevB.76.035439
\bibitem{pm61}
Peelaers H, Hernandez-Nieves AD, Leenaerts O, Partoens B, Peeters FM
Vibrational properties of graphene fluoride and graphane.
Appl. Phys. Lett. 98, 051914. https://doi.org/10.1063/1.3551712
\bibitem{sav1}
Savin AV, Kivshar YuS (2008) 
Discrete breathers in carbon nanotubes.
Europhys. Letters 82, 66002. https://doi.org/10.1209/0295-5075/82/66002
\bibitem{sav2}
Gunlycke D, Lawler HM, White CT (2008)
Lattice vibrations in single-wall carbon nanotubes.
Phys. Rev B 77, 014303. https://doi.org/10.1103/PhysRevB.77.014303

\end{references}
\end{document}